\pdfoutput=1

\documentclass[a4paper,11pt]{article}
\usepackage{jcappub}
\usepackage[T1]{fontenc}
\usepackage{microtype}
\usepackage{graphicx}

\newcommand{\bb}{\ensuremath{\beta\beta}}
\newcommand{\bbonu}{\ensuremath{\beta\beta0\nu}}
\newcommand{\bbtnu}{\ensuremath{\beta\beta2\nu}}

\newcommand{\mbb}{\ensuremath{m_{\beta\beta}}}

\newcommand{\ckky}{\ensuremath{\rm counts/(keV \cdot kg \cdot y)}}

\newcommand{\Qbb}{\ensuremath{Q_{\beta\beta}}}

\newcommand{\AG}{\ensuremath{^{110m}}Ag}

\newcommand{\ND}{\ensuremath{{}^{150}{\rm Nd}}}
\newcommand{\XE}{\ensuremath{{}^{136}\rm Xe}}

\begin{document}

\title{Discovery potential  of xenon-based neutrinoless double beta decay experiments in light of small angular scale CMB observations}

\author{J.J.~G\'omez-Cadenas,} 
\author{J.~Mart\'in-Albo,}
\author{J.~Mu\~noz Vidal}
\author{and C.~Pe\~na-Garay}
\affiliation{Instituto de F\'isica Corpuscular (IFIC), CSIC \& Universitat de Valencia\\
Calle Catedr\'atico Jos\'e Beltr\'an, 2, 46090 Paterna, Valencia, Spain}

\emailAdd{gomez@mail.cern.ch}
\emailAdd{jmalbos@ific.uv.es}
\emailAdd{jmunoz@ific.uv.es}
\emailAdd{penya@ific.uv.es}

\abstract{The South Pole Telescope (SPT) has probed an expanded angular range of the CMB temperature power spectrum. Their recent analysis of 
the latest cosmological data prefers nonzero neutrino masses, $\sum m_\nu = 0.32 \pm 0.11$ eV. This result, if confirmed by the upcoming 
Planck data, has deep implications on the discovery of the nature of neutrinos. In particular, the values of the effective neutrino mass \mbb\ involved in neutrinoless double beta decay (\bbonu) are severely constrained for both the direct and inverse hierarchy, making a discovery much more likely. 
In this paper, we focus in xenon-based \bbonu\ experiments, on the double grounds of their good performance and the suitability of the technology to large-mass scaling. We show that the current generation, with effective masses in the range of 100 kg and conceivable exposures in the range of 500 kg$\cdot$year, could already have a sizable opportunity to observe \bbonu\ events, and their combined discovery potential is quite large. The next generation, with an exposure in the range of 10 ton$\cdot$year, would have a much more enhanced sensitivity, in particular due to the very low specific background that all the xenon technologies (liquid xenon, high-pressure xenon and xenon dissolved in liquid scintillator) can achieve. In addition, a high-pressure xenon gas TPC also features superb energy resolution. We show that such detector can fully explore the range of allowed effective Majorana masses, thus making a discovery very likely. }

\maketitle

\section{Introduction} \label{sec:intro}
Neutrinos, unlike the other Standard Model fermions, could be Majorana particles, that is, indistinguishable from their antiparticles. The existence of Majorana neutrinos would have profound implications in particle physics and cosmology. If neutrinos are Majorana particles, there must exist a new scale of physics, the level of which is inversely proportional to neutrino masses, that characterises new underlying dynamics beyond the Standard Model. The existence of such a new scale provides the simplest explanation of why neutrino masses are so much lighter than the charged fermions. Understanding the new physics that underlies neutrino masses is one of the most important open questions in particle physics. It could have profound implications in our understanding of the mechanism of symmetry breaking, the origin of mass and the flavour problem \cite{Hernandez:2010mi}.

Furthermore, the existence of Majorana neutrinos would imply that lepton number is not a conserved quantum number which could be the origin of the matter-antimatter asymmetry observed in the Universe. The new physics related to neutrino masses could provide a new mechanism to generate the asymmetry, called leptogenesis. Although the predictions are model dependent, two essential ingredients must be confirmed experimentally: 1) the violation of lepton number and 2) CP violation in the lepton sector.

The only practical way to establish experimentally that neutrinos are their own antiparticle is the detection of neutrinoless double beta decay (\bbonu). This is a postulated very slow radioactive process in which a nucleus with $Z$ protons decays into a nucleus with $Z+2$ protons and the same mass number $A$, emitting two electrons that carry essentially all the energy released (\Qbb). The process can occur if and only if neutrinos are massive, Majorana particles.

Several underlying mechanisms --- involving, in general, physics beyond the Standard Model  --- have been proposed for \bbonu, the simplest one being the virtual exchange of light Majorana neutrinos. Assuming this to be the dominant process at low energies, i.e., there are only three light neutrino mass eigenstates, the half-life of \bbonu\ can be written as
\begin{equation}
(T^{0\nu}_{1/2})^{-1} = G^{0\nu} \ \big|M^{0\nu}\big|^{2} \ \mbb^{2} \, .
\label{eq:Tonu}
\end{equation}
In this equation, $G^{0\nu}$ is an exactly-calculable phase-space integral for the emission of two electrons; $M^{0\nu}$ is the nuclear matrix element (NME) of the transition, which has to be evaluated theoretically; and \mbb\ is the \emph{effective Majorana mass} of the electron neutrino:
\begin{eqnarray}
\mbb &=& \Big| \sum_{i} U^{2}_{ei} \ m_{i} \Big| =\nonumber \\
& & \Big| |U_{e1}|^2 m_1 + |U_{e2}|^2 m_2 e^{i\alpha_1} + |U_{e3}|^2 m_3 e^{i\alpha_2}  \Big| 
\label{eq:mbb}
\end{eqnarray}
where $m_{i}$ are the neutrino mass eigenstates and $U_{ei}$ are elements of the neutrino mixing matrix. 

The matrix elements $U_{ei}$~have been stablished by neutrino oscillation experiments, which also measure the mass differences $\delta m^2 = m_2^2 - m_1^2$~and $\Delta m^2 = m_3^2 - m_2^2$. Instead, the two Majorana phases $\alpha_1$ and $\alpha_2$ are unknown. 

On the other hand, cosmological observations, probe the sum of the three neutrino masses:
\begin{equation}
\sum m_\nu= m_1 + m_2 +m_3
\label{eq:cosmo}
\end{equation}

Combining equations~(\ref{eq:mbb}) and (\ref{eq:cosmo}) one can solve for the individual values of the masses, by imposing an additional constrain, namely $\Delta m^2 >0$~(the so-called
``normal hierarchy'') or $\Delta m^2 <0$~(the so-called
``inverse hierarchy'').

The analysis of recent cosmological observations, including the South Pole Telescope (SPT) observations, do indeed show the evidence 
of a finite value for the neutrino cosmological mass, $\sum m_\nu= 0.32 \pm 0.11$ \cite{Hou:2012xq}. This important result does not seem to be 
verified by the analysis including the new Atacama Cosmology Telescope (ACT) observations \cite{ACT}, which nevertheless correspond to a much 
smaller sky coverage than the SPT ones. The cosmological analysis including Planck \cite{Planck} results, coming in a few months, will clarify the evidence of nonzero neutrino cosmological mass claimed by the SPT team.  In this paper we explore how this result affects the discovery 
potential of \bbonu\ experiments.

In previous works  \cite{GomezCadenas:2011it, GomezCadenas:2010gs}, we have made detailed comparisons between the discovery potential of all the major \bbonu\ experiments in the field. In this paper we focus exclusively in xenon-based experiments. Our main argument for doing so is the fact that \XE\ is, by far, the cheapest \bbonu\ decaying isotope, so much so, that a ton of enriched xenon is already available in the world. In addition, the best current sensitivity to \mbb\ is obtained by two xenon experiments: EXO-200  \cite{Auger:2012ar}, a liquid xenon TPC, and KamLAND-Zen  \cite{Gando:2012zm}, a large calorimeter where the xenon is dissolved in liquid scintillator. The combination of both results, all but excludes the long-standing claim of a positive observation  \cite{KlapdorKleingrothaus:2001ke,Gando:2012zm,Bergstrom:2012nx}. Furthermore, recent results from the NEXT
experiment  \cite{Alvarez:2012haa, Alvarez:2012as, Alvarez:2012hh, Alvarez:2012nd}, a high-pressure xenon gas TPC with electroluminescent readout  \cite{Nygren:2009zz,GomezCadenas:2012jv} show excellent resolution and a very low expected background rate, due to the availability of a topological signature (the observation of the two electrons emitted in the decay) which allows a powerful discrimination between signal and background.   

This work is organised as follows: we first summarise the measurement of the total neutrino mass inferred by the analysis of the 
latest cosmological data and then explore the implication of this result on the predictions of the neutrinoless 
double beta effective mass. Next, we discuss the current generation of neutrinoless double beta decay experiments and concentrate particularly on 
the xenon-based experiments, KamLAND-Zen, EXO and NEXT. Finally, we discuss the sensitivity of current and future ton-scale xenon experiments 
to the effective mass and conclude with the impact of  the total neutrino mass measurement in the vicinity of 300 meV on the resolution of 
the crucial question in neutrino physics by xenon--based experiments: are neutrinos their own antiparticles?

\section{Cosmological observations and the neutrino mass}
Cosmological observations can test the sum of the neutrino masses ($\sum m_\nu$), due to the impact of these on the rate of expansion 
and on the growth of perturbations. In fact, a tight upper bound of about 0.3 eV (95 \% CL) 
is derived when the analysis includes cosmic microwave background (CMB)  \cite{WMAP}, baryonic acoustic 
oscillations (BAO)  \cite{BAO1,BAO2,BAO3} and Hubble constant (H$_0$)  \cite{H01,H02} data combined with either 
abundances of Sunyaev-Zeldovich selected galaxy clusters  \cite{clusters} or galaxy clustering data  \cite{Giusarma1,Giusarma2}.

Measurements of the damping tail of the CMB, through the effect of gravitational lensing, are sensitive to the 
low-redshift universe and break the geometric degeneracy present in the large-scale CMB data. The South Pole 
Telescope (SPT) has probed an expanded angular range of the CMB temperature power spectrum and 
confirms a trend for a decreasing power at high multipoles  \cite{Story} relative to the expectation of the 
$\Lambda$CDM model determined by the CMB data at lower multipoles  \cite{Bond:1997wr,Zaldarriaga:1997ch}. This trend can be 
accommodated by a scale-dependent tilt that becomes increasingly red at higher multipoles. Cosmological 
data can not single out the extension of the model needed to accommodate the data. In particular, nonzero neutrino 
masses, smaller helium abundance than predicted by Big Bang nucleosynthesis, running of the scalar spectral index, 
extra relativistic species or nonzero early dark energy (and possible combinations of these) are extensions that could 
explain the combined set of data. 
 
Let us concentrate on the sensitivity of cosmological data to the total neutrino mass, which is known to be larger than $\sim$ 58 meV
 by neutrino oscillation data  \cite{GonzalezGarcia}. The combined analysis of CMB, BAO, H$_0$ data and Sunyaev-Zeldovich 
 selected galaxy clusters abundances prefer nonzero neutrino masses, $\sum m_\nu = 0.32 \pm 0.11$ eV \cite{Hou:2012xq}. The 
 significant improvement in the CMB data by SPT leads to a better determination of the spectral index, which is  
 anticorrelated with the total mass of neutrinos. The addition of the other probes, particularly BAO and cluster abundances, further 
 improve the constraints. Other cosmological parameters may decrease the significance (spatial curvature, running of the scalar 
 spectral index), increase the significance (equation of state parameter of dark energy and effective number of neutrino species) 
 or be insensitive (Helium abundance). In fact, the model with extra effective neutrinos (N$_{eff}$) and nonzero total neutrino mass 
 is the best model of the CMB+BAO+H$_0$ data.  The combined analysis of CMB, BAO, H$_0$ data and Sunyaev-Zeldovich 
 selected galaxy clusters abundances prefer nonzero neutrino masses, $\sum m_\nu = 0.51 \pm 0.15$ eV  
 and N$_{eff} = 3.86 \pm 0.37$  \cite{Hou:2012xq}. In this work, we explore the standard scenario of three light neutrinos and 
 defer to a future work the more exotic possibility of extra sterile neutrino states.

Recently, the ACT team has presented their analysis of cosmological observations including the small angular scale 
CMB observations by the Atacama telescope  \cite{ACT}. Their observations are consistent with a zero neutrino mass and correctly 
point to the fact that the nonzero mass determined by the SPT team requires a rather low amplitude of the linear power spectrum 
on the scale of 8 Mpc/h, $\sigma_8$, which is in tension with their cluster and skewness measurements. 
Nevertheless, SPT has a lot more sky coverage. Certainly, the analysis of these recent observations needs more discussion and 
will be further tested by including observations of the upcoming cosmological Planck CMB data  \cite{Planck}.

\section{Neutrinoless double beta rate derived by neutrino data} \label{sec:rate}
The determination of the total mass of neutrinos has an important impact on a crucial question in neutrino physics, i.e, whether 
the neutrino is a Dirac or a Majorana fermion. The question is resolved if neutrinoless double beta decay is observed. On the 
other hand, long lifetimes may not be accessible experimentally and the question would remain unsolved. We will show next, that the 
total neutrino mass derived by cosmological data \cite{Hou:2012xq} leads to an upper bound on the lifetime, which can be reached experimentally, in particular by (multi)ton xenon-based experiments. 

The neutrinoless double beta decay rate is proportional to the  effective mass \mbb\, equation~({eq:Tonu}) , which is 
given by the sum of three terms which may have partial cancellation among them, equation~(\ref{eq:mbb}). The total neutrino mass 
derived by cosmological observations with the mass squared differences measured by reactor neutrino experiments determine the 
masses of the free neutrinos m$_i$. The moduli of the mixing matrix elements are well measured by solar and reactor neutrinos experiments, 
where $|U_{e1}|=\cos \theta_{12} \cos \theta_{13}$, $|U_{e2}|=\sin \theta_{12}\cos \theta_{13}$ and $|U_{e3}|=\sin \theta_{13}$. The relative 
phases between the three terms are free unknown parameters. In the set of measured neutrino parameters, the total mass of neutrinos 
is the most uncertain.  

\begin{figure}
\begin{center}
\includegraphics[width=15.cm]{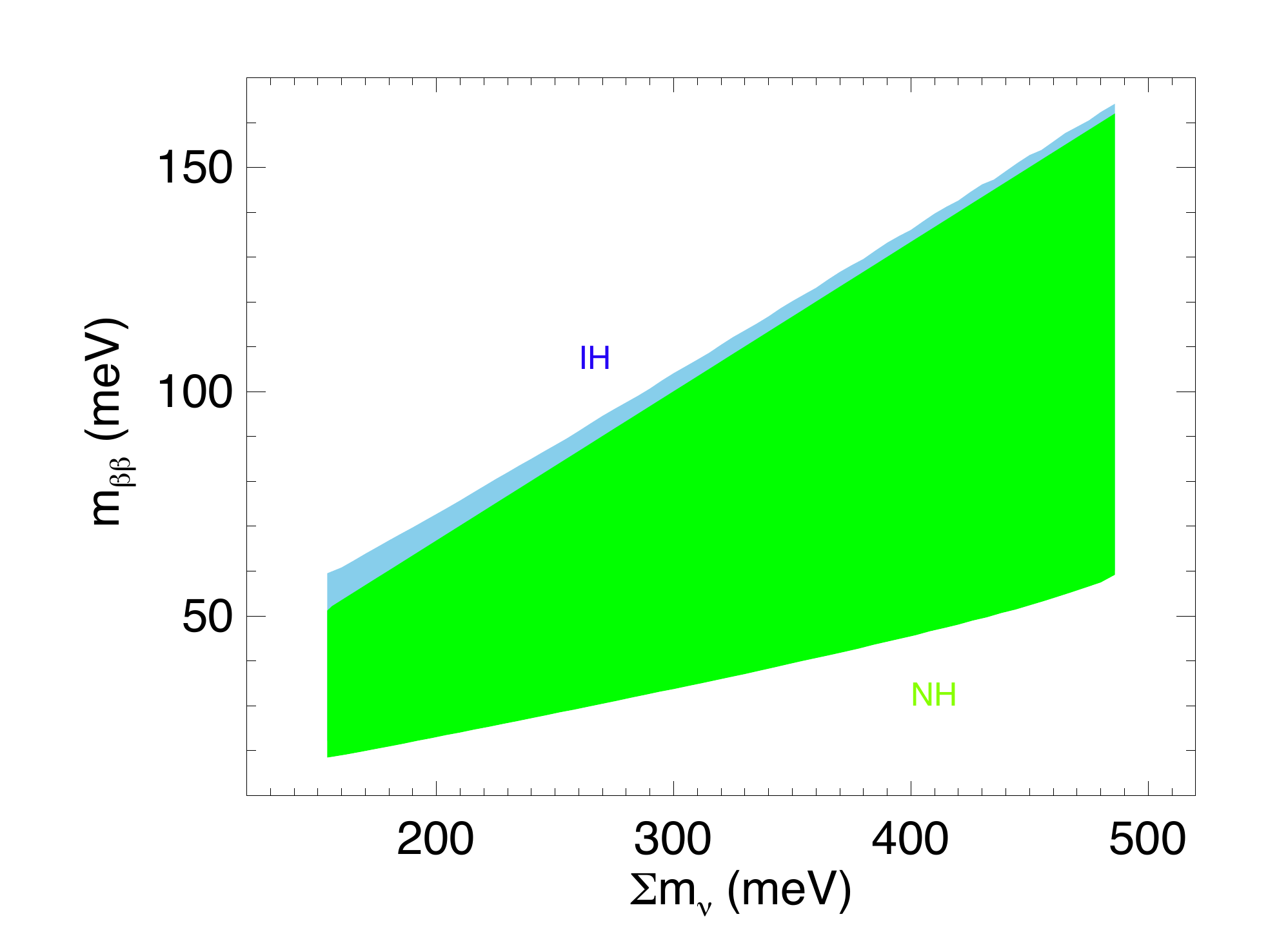}
\caption{1$\sigma$ allowed regions for two degrees of freedom in the observables space \mbb\  and $\sum m_\nu$ by 
neutrino oscillation and cosmological data, assuming normal (green) and inverted (sky blue) mass hierarchy.}
\label{mee_vs_sum}
\end{center}
\end{figure} 

We have explored the predictions of  \mbb\ derived by the measured neutrino oscillations 
parameters shown in the most up-to-date analysis  \cite{GonzalezGarcia}: $\sin^2 \theta_{12}= 0.302  \pm 0.013$, 
$\sin^2 \theta_{13}= 0.023 \pm 0.002$, $\delta m^2=75 \pm 2$ meV$^2$,  
$\Delta m^2=2470 \pm 70$ meV$^2$ ($\Delta m^2=-2427^{+42}_{-65}$ meV$^2$)
for the normal (inverted) neutrino mass hierarchy.

Figure \ref{mee_vs_sum} shows the 1$\sigma$ allowed region for two degrees of freedom (dof) in the observables 
space \cite{GonzalezGarcia:2010un},  \mbb\  and $\sum m_\nu$, both in meV.  
The regions are calculated with the assumption of  gaussian uncorrelated errors in the neutrino parameters derived by the global analysis 
of neutrino oscillation and cosmological data, and varying the Majorana phases within their physical range. The two regions correspond to the normal 
and inverted mass hierarchy scenarios. We can see that the mass determined by cosmological observations, large compared to the mass 
splittings, leads to a quasi-degenerate spectrum. Normal and inverted hierarchy lead to similar neutrinoless double beta rate predictions. 
The effective mass \mbb\  is smaller than a maximum value close to a third of the sum of neutrino masses, as expected in a degenerate spectrum.  
More importantly, \mbb\  is also  bound  from below. The variation of the unknown Majorana phases can not completely cancel the effective mass, 
which is larger than $\sim$ 20 meV. 
The 1$\sigma$ allowed range of the \mbb\ parameter by present neutrino oscillation and cosmological data is
\begin{itemize}
\item $26 \le \mbb \le 143$ for the normal hierarchy;
\item $28 \le \mbb \le 145$ for the inverted hierarchy.
\end{itemize}

We find that, irrespectively of the (quasi-degenerated) hierarchy, the 1$\sigma$ range of \mbb\ is $[26,145]$. 
Therefore, a confirmation of  
the results presented in  \cite{Hou:2012xq} are very important to validate the 20-meV target sensitivity for neutrinoless double beta decay 
experiments, needed to identify the nature of neutrinos irrespectively of the mass hierarchy.

\section{The current generation of \boldmath{\bbonu} experiments} \label{sec:current}
The detectors used to search for \bbonu\ are designed, in general, to measure the energy of the radiation emitted by a \bbonu\ source. In a neutrinoless double beta decay, the sum of the kinetic energies of the two released electrons is always the same, and equal to the mass difference between the parent and the daughter nuclei: $\Qbb \equiv M(Z,A)-M(Z+2,A)$. However, due to the finite energy resolution of any detector, \bbonu\ events would be reconstructed within a given energy range centred around \Qbb\ and typically following a gaussian distribution. Other processes occurring in the detector can fall in that region of energies, thus becoming a background and compromising drastically the sensitivity of the experiment  \cite{GomezCadenas:2010gs}.

All double beta decay experiments have to deal with an intrinsic background, the standard two-neutrino double beta decay (\bbtnu), that can only be suppressed by means of good energy resolution. Backgrounds of cosmogenic origin force the underground operation of the detectors. Natural radioactivity emanating from the detector materials and surroundings can easily overwhelm the signal peak, and hence careful selection of radiopure materials is essential. Additional experimental signatures, such as event topological information, that allow the distinction of signal and background are a bonus to provide a robust result.

Besides energy resolution and control of backgrounds, several other factors such as detection efficiency and scalability to large masses must be taken into consideration in the design of a double beta decay experiment. The simultaneous optimisation of all these parameters is most of the time conflicting, if not impossible, and consequently many different experimental techniques have been proposed. In order to compare them, a figure of merit, the experimental sensitivity to \mbb, is normally used  \cite{GomezCadenas:2010gs}:
\begin{equation}
\mbb \propto \sqrt{1/\varepsilon}\, \left(\frac{b\ \delta E}{M\ t} \right)^{1/4}, \label{eq:sensi}
\end{equation}
where $\varepsilon$ is the signal detection efficiency, $M$ is the \bb\ isotope mass used in the experiment, $t$ is the data-taking time, $\delta E$ is the energy resolution and $b$ is the background rate in the region of interest around \Qbb\ (expressed in counts per kg of \bb\ isotope, year and keV, henceforth abbreviated as {\em ckky}).

 The status of the field has been the subject of several recent reviews  \cite{GomezCadenas:2011it, Cremonesi:2012av, Sarazin:2012ct, Zuber:2006hv}. Among the on-going and planned experiments, many different experimental techniques are utilised, each with its pros and cons. The time--honored approach of emphasising energy resolution and detection efficiency is currently spearhead by germanium calorimeters like GERDA  \cite{Cattadori:2012fy}  and {\sc Majorana} \cite{Wilkerson:2012ga}, as well as tellurium bolometers such as CUORE \cite{Gorla:2012gd}. 
 
 A different, and powerful approach, the main topic of this paper, proposes the use of xenon-based experiments. Two of them, EXO-200 \cite{Auger:2012gs} and KamLAND-Zen \cite{Gando:2012jr} are already operating, while NEXT \cite{Alvarez:2012haa} is in the initial stages of construction, and foresees to start taking data in 2015.

Other experiments that will operate in the next few years are the SuperNEMO demonstrator \cite{Sarazin:2012ct}, a tracker-calorimeter approach which provides a powerful topological signal (the observation of the two electrons emitted in a \bb\ decay) but is hard to extrapolate to larger masses (the demonstrator itself will have a mass of less than 10 kg of isotope), and SNO+ \cite{GomezCadenas:2011it}, a large liquid scintillator calorimeter (the same approach that KamLAND-Zen), in which natural Neodymium is dissolved in the scintillator. Neodymium is a very interesting isotope, whose large \Qbb\ suppresses many of the low-energy background than other experiments have to deal with, but the \bbonu\ decaying isotope, \ND\ is only 5\% of the natural  Neodymium, limiting the total mass that the experiment can deploy.

\section{Xenon experiments} \label{sec:xenon}
Xenon is an almost-optimal element for \bbonu\ searches, featuring many desirable properties, both as a source and as a detector. It has two naturally occurring-isotopes that can decay via the \bb\ process, $^{134}$Xe ($\Qbb=825$~keV) and \XE\ ($\Qbb=2458$~keV). The latter, having a higher $Q$ value, is preferred since the decay rate is proportional to $\Qbb^{5}$ and the radioactive backgrounds are less abundant at higher energies. Moreover, the \bbtnu\ mode of \XE\ is slow ($\sim2.3\times10^{21}$~years \cite{KamLANDZen:2012aa,Ackerman:2011gz}) and hence the experimental requirement for good energy resolution to suppress this particular background is less stringent than for other \bb\ sources. The process of isotopic enrichment in the isotope \XE\ is relatively simple and cheap compared to that of other \bb\ isotopes. Xenon has no long-lived radioactive isotopes and is therefore intrinsically clean.

As a detector, xenon is a noble gas, and therefore one can build a time projection chamber (TPC) with pure xenon as detection medium. Both a liquid xenon (LXe) TPC and a (high-pressure) gas (HPXe) TPC are suitable technologies, chosen by the EXO-200 and the NEXT experiment respectively. Being a noble gas, xenon can also be dissolved in liquid scintillator. This is the approach of the KamLAND-Zen experiment. 



\subsection{KamLAND-Zen}
The KamLAND-Zen experiment is a modification of the well-known KamLAND neutrino detector  \cite{Abe:2009aa}. A transparent balloon, $\sim3$~m diameter, containing 13 tons of liquid scintillator loaded with 320 kg of xenon (enriched to 91\% in \XE) is suspended at the centre of KamLAND. The scintillation light generated by events occurring in the detector is recorded by an array of photomultipliers surrounding it. From the detected light pattern, the position of the event vertex is reconstructed with a spatial resolution of about $15~\mathrm{cm}/\sqrt{E(\mathrm{MeV})}$. The energy resolution is $(6.6\pm0.3)\%/\sqrt{E(\mathrm{MeV})}$, that is, 9.9\% FWHM at the $Q$ value of \XE. The signal detection efficiency is $\sim0.42$ due to the tight fiducial cut introduced to reject backgrounds originating in the balloon. The achieved background rate in the energy window between 2.2~MeV and 3.0~MeV is $10^{-3}$~\ckky. 

KamLAND-Zen
has searched for \bbonu\ events with an exposure of 89.5 kg$\cdot$year. They have published a limit on the half-life of \bbonu\ of $T_{1/2}^{0\nu}(\XE) > 1.9 \times 10^{25}$ years \cite{Gando:2012zm}. 

\subsection{EXO}

The EXO-200 detector \cite{Auger:2012gs} is a symmetric LXe TPC deploying 110 kg of xenon (enriched to 80.6\% in \XE).

In EXO-200, ionisation charges created in the xenon by charged particles drift under the influence of an electric field towards the two ends of the chamber. There, the charge is collected by a pair of crossed wire planes which measure its amplitude and transverse coordinates. Each end of the chamber includes also an array of avalanche photodiodes (APDs) to detect the 178-nm scintillation light. The sides of the chamber are covered with teflon sheets that act as VUV reflectors, improving the light collection. The simultaneous measurement of both the ionisation charge and scintillation light of the event may in principle allow to reach a detector energy resolution as low as 3.3\% FWHM at the \XE\ Q-value, for a sufficiently intense drift electric field \cite{Conti:2003av}. 

The EXO-200 detector achieves currently an energy resolution of 4\% FWHM at \Qbb, and a background rate measured in the \emph{region of interest} (ROI) of $ 1.5 \times 10^{-3}\ckky$. The experiment has also searched for \bbonu\ events. The total exposure used for the published result is 32.5 kg$\cdot$year. They have published a limit on the half-life of \bbonu\ of $T_{1/2}^{0\nu}(\XE) > 1.6 \times 10^{25}$ years \cite{Auger:2012ar}.

The combination of the KamLAND-Zen and EXO results yields a limit $T_{1/2}^{0\nu}(\XE) > 3.4 \times 10^{25}$ years, which essentially excludes the long-standing claim of Klapdor-Kleingrothaus and collaborators \cite{KlapdorKleingrothaus:2001ke}  \cite{{Gando:2012zm},Bergstrom:2012nx}.

\subsection{NEXT}

The NEXT experiment \cite{Alvarez:2012haa} will search for the neutrinoless double beta decay of \XE\ using an asymmetric high pressure gas xenon (HPXe) TPC, filled with 100--150 kg of  xenon (enriched to 91\% in \XE) gas at 15--20 bar pressure. NEXT 
offers two major advantages for the search of neutrinoless double beta decay, namely: a) {\em excellent energy resolution}, with an intrinsic limit of about 0.3\% FWHM at \Qbb\ and 0.5--0.7\% demonstrated by the large-scale prototypes NEXT-DBDM and NEXT-DEMO \cite{Alvarez:2012hh, Alvarez:2012nd}, b) {\em tracking capabilities} that provide a powerful topological signature to discriminate between signal (two electron tracks with a common vertex) and background (mostly, single electrons). The topological signature, combined with a radio clean detector results in a very low specific background rate.

The combination of radio purity and the additional rejection power provided by the topological signature of the two electrons results in an expected background rate of $10^{-4} -- 5 \times 10^{-4}$ \ckky, depending of the level of background of the energy plane PMTs. There are only upper limits for those PMTs. The most sensitive measurement, performed by the LUX collaboration, quotes am upper limit in the background of each PMT of less than 700 $\mu$Bq, and corresponds to the lowest limit of the background rate, while the XENON collaboration quotes a less sensitive limit that results in the upper limit of the background rate. The NEXT collaboration is currently screening all the PMTs entering the detector energy plane. While the measurement program is going on, they quote the upper limit of their background level, $5 \times 10^{-4}$ \ckky, as reference  \cite{Alvarez:2012haa, MartinAlbo:2013xx}. The construction of the detector is underway at the Laboratorio Subterr\'aneo de Canfranc (LSC), in Spain. NEXT owns 100 kg of enriched xenon, and foresees to start a physics run in 2015.

\section{Sensitivity of xenon experiments} \label{sec:sensi}

\subsection{Sensitivity of the current xenon experiments}

\subsubsection{Experimental parameters}

\begin{table}
\centering
\caption{Experimental parameters of the three xenon-based double beta decay experiments: (a) total mass of \XE, $M$; (b) enrichment fraction $f$; (c) signal detection efficiency, $\varepsilon$; (d) energy resolution, $\delta E$, at the $Q$ value of \XE; and background rate, $b$, in the region of interest around \Qbb\ expressed in \ckky\ (shortened as ckky). } 
\label{tab:ExpParams}
\vspace{0.5cm}
\begin{tabular}{lccccc}
\hline
Experiment & $M$ (kg) & $f$ (\%) & $\varepsilon$ (\%) & $\delta E$ (\% FWHM) & $b$ ($10^{-3}$~ckky) \\
\hline
EXO-200 		& 110 & 0.81& 0.56 & 4.0  & 1.5 \\
KamLAND-Zen & 330 & 0.91 & 0.42 & 9.9 & 1.0  \\
NEXT-100 	& 100 & 0.91 & 0.30  & 0.7  & 0.5 \\
\hline
\end{tabular}
\end{table}

The experimental parameters of the three xenon experiments described here, as defined in equation~(\ref{eq:sensi}), are collected in Table~\ref{tab:ExpParams}. The parameters of EXO-200 and KamLAND-Zen are those published by the collaborations \cite{Auger:2012ar,Gando:2012zm}. The resolution in NEXT corresponds to the most conservative result obtained by their prototypes  \cite{Alvarez:2012nd}, and the predicted background rate and efficiency comes from the full background model of the collaboration \cite{Alvarez:2012haa, MartinAlbo:2013xx}, assuming a conservative background level for the dominant source of background (the energy--plane PMTs, see discussion in the previous section). 

A caveat is in order concerning NEXT. Although the resolution is solidly established by the NEXT-DEMO and NEXT-DBDM prototypes, and the different components that will enter the detector have been carefully screened \cite{Alvarez:2012as}, to construct the background model, the predictions of the Monte Carlo have not been validated with actual data from the operating detector, as the other two experiments have already done. In this sense, the comparisons in this work are intended to show the potential of the technology, rather than its demonstrated performance. This said, the availability of a topological signature (also clearly established by the prototypes), the excellent resolution, and the on--going campaign to screen every component that goes into the detector, builds a strong case in favour of the HPXe technology. 

\subsubsection{Nuclear matrix elements}

In order to compute a sensitivity plot, one has to choose a given set of nuclear matrix elements (NME). In the last few years the reliability of the calculations has been addressed, with several techniques being used (see  \cite{GomezCadenas:2010gs} and references therein), namely: the Interacting Shell Model (ISM); the Quasiparticle Random Phase Approximation; the Interacting Boson Model (IBM); and the Generating Coordinate Method (GCM). In most cases the results of the ISM calculations are the smallest ones, while the largest ones come often from IBM.

Each one of the major methods has some advantages and drawbacks. The clear advantage of the ISM calculations is their full treatment of the nuclear correlations, while their drawback is that they may underestimate the NMEs due to the limited number of orbits in the affordable valence spaces. It has been estimated  \cite{GomezCadenas:2010gs} that this effect can be of the order of 25\%. On the contrary, the QRPA variants, the GCM and the IBM are bound to underestimate the multipole correlations in one or another way. As it is well established that these correlations tend to diminish the NMEs, these methods should tend to overestimate them.

With this considerations in mind, a physics-motivated range (PMR) of theoretical values for the NMEs of different isotopes was proposed in  \cite{GomezCadenas:2010gs}. In the case of \XE\, the PMR range extends from a lower limit, defined by the ISM model, with a NME of 2.2 to an upper limit, defined by the IBM model, with a NME of $\sim$4. The central PMR value, used for all plots in this paper is 2.9. See  \cite{GomezCadenas:2010gs} for further discussion.

\subsubsection{Sensitivity}

\begin{figure}
\centering
\includegraphics[width=0.7\textwidth]{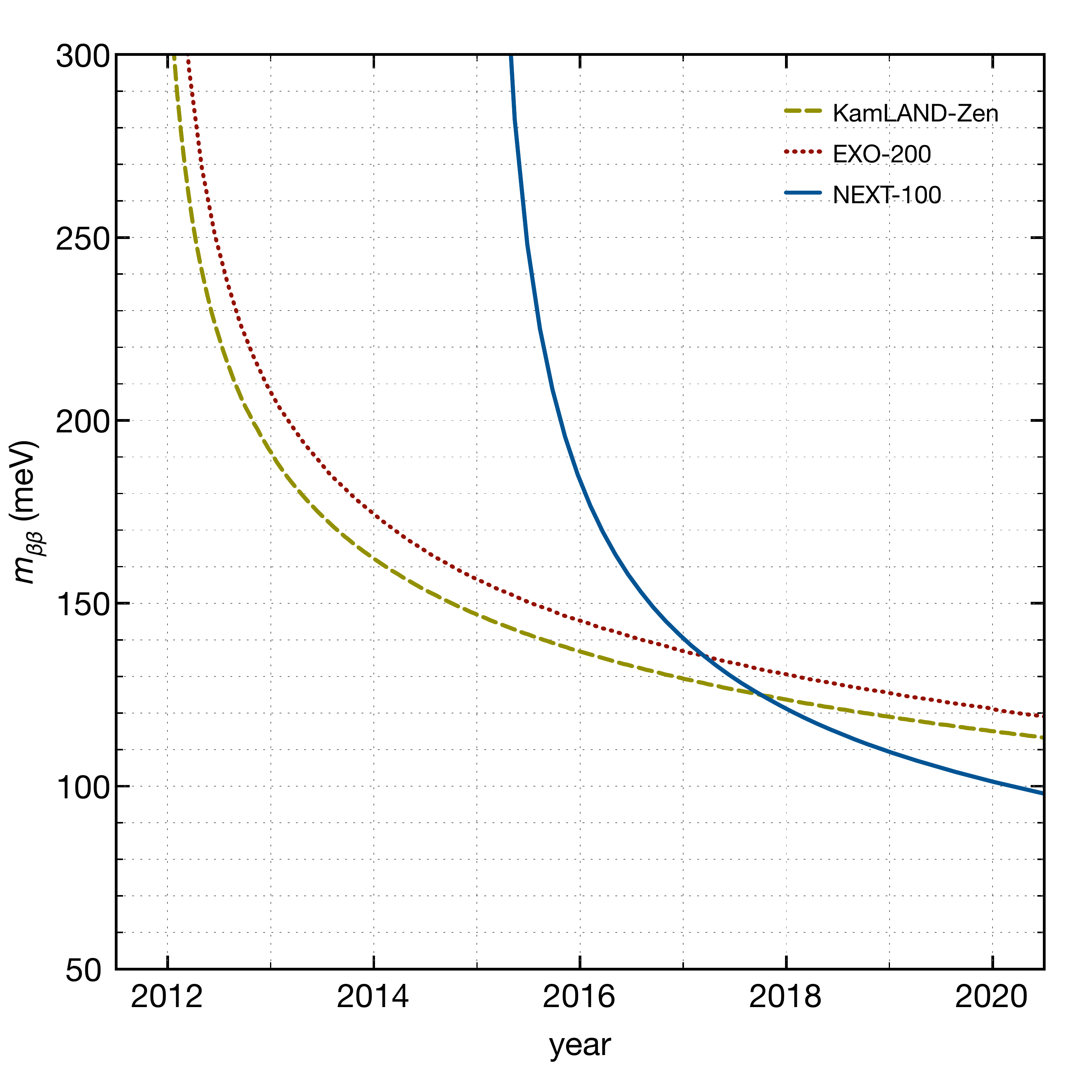}
\caption{Sensitivity of the three xenon experiments as a function of the running time, assuming the parameters described 
in Table \ref{tab:ExpParams}. We consider a run of 8 years for EXO-200 and KamLAND-Zen (2012 to 2020) and a run of 5 years for NEXT (2015 to 2020).} \label{fig.exoNext}
\end{figure}

Figure \ref{fig.exoNext}, shows the expected performance of the three experiments, assuming
the parameters described 
in Table \ref{tab:ExpParams} and the central value of the PMR described above. 
We consider a run of five years for NEXT (2015 to 2020) and a longer run of eight years for EXO-200 and KamLAND-Zen (2012 to 2020). A total dead-time of 10\% a year for all experiments is assumed. 
Observe the following features:

\begin{itemize}
\item Most of the gains occur in the first two years. Once the experiments enter in the regime of being background dominated, progress is very slow, as predicted by equation~(\ref{eq:sensi}).
\item The curve corresponding to the NEXT experiment drops faster than that corresponding to the other two experiments, due to better energy resolution and background suppression. This compensates its late start. 
\item By 2018 (6 years run in the case of EXO-200 and KamLAND-ZEN, 3 years run in the case of NEXT), all the experiment reach a similar sensitivity of about 130 meV. 
\item By 2020, the NEXT experiment reaches 103 meV, KamLAND-ZEN reaches 115 meV and EXO
123 meV. 
\end{itemize}

It follows that all the three experiment will have a chance of making a discovery if \mbb\ is in the upper part of its allowed range, see Figure \ref{mee_vs_sum}. The fact that the experiments are based in different experimental techniques, with different systematic errors makes their 
simultaneous running even more attractive. The combination of the three can reach a sensitivity of about 65 meV, which covers a significant 
fraction of the phase space. This result is affected by uncertainties in the values of the NME. Taking the lower bound of the PMR we find a sensitivity 
of 87 meV for the combined limit, while taking the IBM model as upper bound of the PMR we find a sensitivity of 48 meV.  

\subsection{Sensitivity of ton-scale xenon experiments}

To study the projected sensitivity of future xenon experiments, we consider  three hypothetical detectors of the same mass (1 ton) running for the same total exposure (up to 10 ton$\cdot$ year) based in the three technologies discussed above: liquid xenon (LXe), xenon--liquid scintillator (XeSci) and high pressure gas xenon (HPXe).  
Our choice of one ton as the reference mass for these studies is motivated by the following reasons:

\begin{itemize}
\item {\bf Availability of the isotope:} there is already one ton of enriched xenon available in the world (most of it owned by the KamLAND-Zen collaboration), that could be pooled in a future one--ton experiment. The cost of one ton of enriched xenon is (currently) rather modest, about 10--20 M\$, typically a factor ten cheaper than the cost of other enriched materials. 
\item {\bf Scalability of the technology:} Building one--ton xenon detectors appears rather feasible without major modifications to the currently operational technologies. A liquid--scintillator calorimeter would simply dissolve more xenon in the scintillator than KamLAND-Zen, eventually building a larger balloon. Given the high density of LXe, a one--ton detector based on this technology is still a very compact object (e.g, a sphere holding 1 ton of LXe would have a radius of only 42.7 cm). In the case of a HPXe operating at 20 bar, about 10 m$^3$~are needed to hold a fiducial mass of 1 ton of xenon. This corresponds to a cylinder of 1 meter radius by 3 meters long, a large, but not huge TPC.
\end{itemize}

Furthermore, considering the same mass for the three technologies and running their sensitivity as a function of the exposure allows to compare their potential in the same level ground.

\subsubsection{Resolution of one-ton xenon detectors}

For HPXe, the intrinsic limit dictated by the Fano factor in xenon gas is 0.3\% FWHM at \Qbb,  but we consider safer to quote the actual resolution measured by the NEXT-DBDM prototype  \cite{Alvarez:2012hh}, which obtains 0.5 \% FWHM at \Qbb.

For LXe, we use the best projected resolution for the technology, 3.3\% FWHM at \Qbb\  \cite{Conti:2003av}, also near the intrinsic limit in liquid xenon.
 
For SciXe, we consider that a liquid scintillator calorimeter can be upgraded (by adding more PMTs) to improve the energy resolution. As a reference we consider  SNO+ detector, which boasts the best energy resolution of all liquid scintillator calorimeters, 6.5\% FWHM at \Qbb.

\subsubsection{Background rate of one-ton xenon detectors}

For  HPXe we take the best case of the NEXT background model, which predicts an specific background rate of
$10^{-4}$~\ckky\ when using the most sensitive limits measured the energy plane PMTs \cite{MartinAlbo:2013xx}. 

For LXe, the current, very low background rate, achieved by the EXO-200 detector, is obtained with only marginal self-shielding. The reason for that is that EXO-200 is a small apparatus, and leaving part of the LXe as a shield has a large cost in efficiency. The situation, however, improves dramatically for a larger detector. 

For the sake of simplicity, consider an spherical LXe detector, with one ton mass and a radius of 43 cm. Leaving a shell of 10 cm of LXe as a shield reduces the specific background by a factor 1/15, and keeps 43\% of the enriched xenon as the fiducial mass of the experiment. Assuming that the selection efficiency of the future LXe experiment will be similar to that of EXO-200 we find that a
one-ton LXe detector could reach a background rate of $10^{-4}$~\ckky\ with an overall efficiency of 38\%. 

Concerning the liquid scintillator calorimeter, we assume that destilation of the \AG\  will result in about one order of magnitude reduction in the specific background, as discussed in  \cite{Kozlov:2011}. For simplicity, we also consider an specific background rate of $10^{-4}$~\ckky\ and leave the efficiency unchanged with respect to KamLAND-Zen.

\subsubsection{Experimental parameters}

\begin{table}
\centering
\caption{Expected experimental parameters of the three xenon-based double beta decay technologies: (a) signal detection efficiency, $\varepsilon$; (b) energy resolution, $\delta E$, at the $Q$ value of \XE; and background rate, $b$, in the region of interest around \Qbb\ expressed in \ckky. } \label{tab:FutureParams}
\vspace{0.5cm}
\begin{tabular}{lccc}
\hline
Experiment &  $\varepsilon$ (\%) & $\delta E$ (\% FWHM) & $b$ ($10^{-3}$~ckky) \\
\hline
LXe		& 0.38 & 3.2  & 0.1 \\
XeSci       & 0.42 & 6.5 & 0.1  \\
HPXe	& 0.30  & 0.5  & 0.1 \\
\hline
\label{tab:para}
\end{tabular}
\end{table}

Table~\ref{tab:FutureParams} summarises our projections of the experimental parameters for the three technologies. Notice that, while we believe that the parameters displayed in Table~\ref{tab:FutureParams} are reasonable, we are not claiming that they represent any specific design.   We assume a resolution near the practical limit for the three technologies, and use reasonable assumptions to predict their achievable background rate, which turns out to be, both very small and quite similar. 

\subsubsection{Sensitivity}
\begin{figure}
\centering
\includegraphics[width=0.7\textwidth]{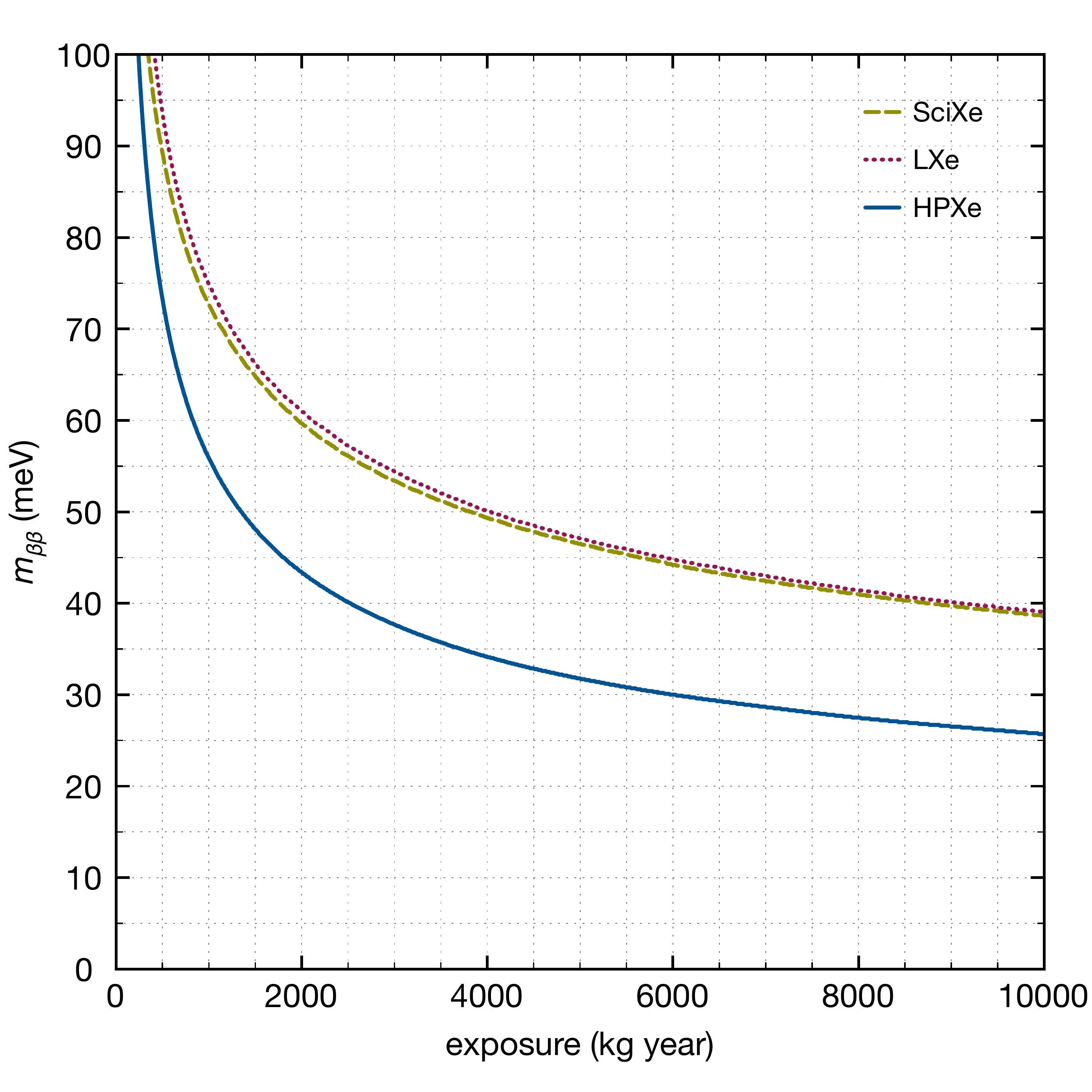}
\caption{Sensitivity of the three technologies experiments as a function of the total exposure, assuming the parameters described 
in Table \ref{tab:FutureParams}.} \label{fig.NEXTW}
\end{figure}

Figure \ref{fig.NEXTW}, shows the expected performance of the three technologies, assuming
the parameters described 
in Table \ref{tab:FutureParams}, up to a total exposure of 10 ton$\cdot$year. Although we have used a reference mass of one ton, the actual detector designs could consider, of course, larger masses. The tradeoff between total detector mass and exposure time needs to be done taking into account detector design and the cost of enriched xenon.

At the maximum exposure, the LXe and and XeSci detectors reach a draw at 40 meV, while the HPXe detector reaches 25 meV. Each one of the experiments covers a large fraction of the available phase space, with HPXe covering practically all the range of allowed values. The combination of the three experiments is 19 meV, fully covering the phase space, while the combination of HPXe and one of the other two is 21 meV. 

This result is, of course, affected by uncertainties in the values of the NME. Taking the lower bound of the PMR we find a sensitivity of 25 meV for the combined limit, while taking the IBM model as upper bound of the PMR we find a sensitivity of 14 meV. Notice that the HPXe (using the central value of the PMR) fully covers the one-sigma range of \mbb\ values ($[26,145]$ meV, see Section \ref{sec:rate}) , while the combination of the three experiments covers the range even for the lower bound of the PMR (that is the ISM, which gives the lowest NME of all the available models). 

\section{Discussion}\label{sec:conclu}

\begin{figure}
\begin{center}
\includegraphics[width=15.cm]{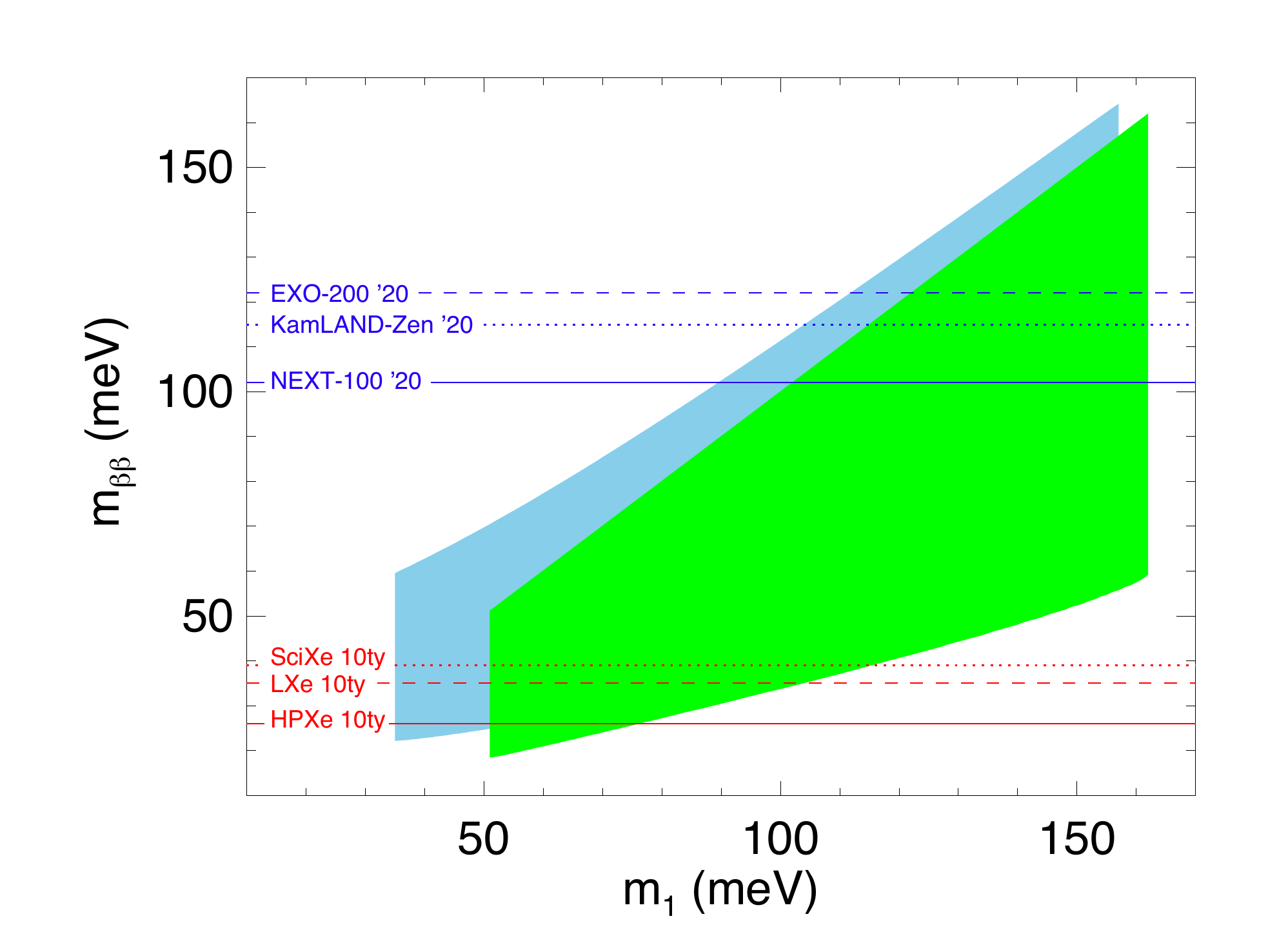}
\caption{Predictions in the parameter space of neutrinoless double beta effective mass and lowest neutrino mass. 
Full regions show the 1$\sigma$ allowed regions (2 dof) by neutrino oscillation and cosmological data, assuming normal (green) and
inverted (sky blue) mass hierarchy. Horizontal lines in blue show the expected sensitivity of current xenon--based 
experiments in 2020. Horizontal lines in red show the expected sensitivity of future xenon--based neutrinoless double beta 
technologies after 10 ton$\cdot$year exposure.}
\label{mee_vs_m1}
\end{center}
\end{figure} 

In this work, we have addressed the question whether present and ton scale xenon--based double beta decay experiments can fully answer 
the quest on the nature of neutrinos or not. We find a positive answer, based on the evidence of a nonzero value for the neutrino cosmological mass, 
$\sum m_\nu= 0.32 \pm 0.11$, determined by the SPT team  \cite{Hou:2012xq} and the assumption of the standard model extended 
by only three light neutrinos. This measured total neutrino mass of light neutrinos implies a quasi--degenerate neutrino mass spectrum, 
what leads to important consequences. 

Our findings are summarized in Figure \ref{mee_vs_m1}, where we use the parameter space of neutrinoless double beta effective mass and lowest 
neutrino mass. The 1$\sigma$ allowed regions by neutrino oscillation and cosmological data show that \mbb\ is bound from below, at the level 
of 20 meV, independently of the neutrino mass hierarchy. In particular, the one-sigma range of \mbb\ is $[26,145]$ meV. The free Majorana 
phases are unable to produce full cancellation of the effective mass due to the degeneracy of the neutrino masses.

We have considered xenon-based \bbonu\ experiments, on the double grounds of their good performance, and the suitability of the technology 
to large-mass scaling. Firstly we discuss the current generation of experiments, KamLAND--Zen, EXO--200 and NEXT, with effective masses in the 
range of 100 kg and conceivable exposures in the range of 500 kg$\cdot$year. The expected sensitivity  
of the three experiments to  \mbb\  in 2020 is shown by blue horizontal lines in Figure \ref{mee_vs_m1}. 
All three experiments have sensitivity to some of the effective mass predicted by neutrino oscillation and total mass measurements, and the 
combination of the three, with sensitivity of 65 meV,  have the potential to test about half of the allowed effective mass parameter space.  The uncertain value of the NME modifies the sensitivity of the combination, from 48 meV to 87 meV. 

More importantly, the lower bound in \mbb\ , implies the potential to distinguish whether neutrinos are Dirac or Majorana particles, under the quoted assumptions. The next generation experiments, with an exposure in the range of 10 ton$\cdot$year, would have a much more enhanced sensitivity, 
as shown by red horizontal lines in Figure \ref{mee_vs_m1}.  Al three technologies have the potential to explore most of the \mbb\ allowed region. 
The high pressure gas xenon TPC, due to the excellent energy resolution, can cover the full range of \mbb\ predictions under reasonable NME assumptions (the central value of the PMR). The combination of the three technologies would cover the full range of allowed \mbb\ values even for the smallest NME.

In summary, xenon experiments may be the tool to demonstrate that neutrinos are Majorana particles in the next few years.  

\acknowledgments
We warmly acknowledge C.~Gonz\'alez-Garc\'ia and P.~Hern\'andez for discussions, help and insight. This work was supported by the \emph{Ministerio de Econom\'ia y Competitividad} of Spain under grants CONSOLIDER-Ingenio 2010 CSD2008-0037 (CUP), FPA2009-13697-C04-04 and FPA2011-29678, and by the Generalitat Valenciana  grant PROMETEO/2009/116 and the ITN INVISIBLES (Marie Curie Actions, PITN-GA-2011-289442).

\bibliographystyle{JHEP}
\bibliography{references}

\providecommand{\href}[2]{#2}\begingroup\raggedright\begin{thebibliography}{10}

\bibitem{Hernandez:2010mi}
P.~Hern\'andez, {\it {Neutrino physics}},
  \href{http://xxx.lanl.gov/abs/1010.4131}{{\tt arXiv:1010.4131}}.

\bibitem{Hou:2012xq}
Z.~Hou et~al., {\it {Constraints on Cosmology from the Cosmic Microwave
  Background Power Spectrum of the 2500-square degree SPT-SZ Survey}},
  \href{http://xxx.lanl.gov/abs/1212.6267}{{\tt arXiv:1212.6267}}.

\bibitem{ACT}
J.~L. Sievers et~al., {\it {The Atacama Cosmology Telescope: Cosmological
  parameters from three seasons of data}},
  \href{http://xxx.lanl.gov/abs/1301.0824}{{\tt arXiv:1301.0824}}.

\bibitem{Planck}
{\bf Planck} Collaboration, P.~A.~R. Ade et~al., {\it {Planck Early Results. I.
  The Planck mission}},  {\em Astron.\ Astrophys.} {\bf 536} (2011) 16464,
  [\href{http://xxx.lanl.gov/abs/1101.2022}{{\tt arXiv:1101.2022}}].

\bibitem{GomezCadenas:2011it}
J.~J. G\'omez-Cadenas, J.~Martin-Albo, M.~Mezzetto, F.~Monrabal, and M.~Sorel,
  {\it {The search for neutrinoless double beta decay}},  {\em Riv.\ Nuovo
  Cim.} {\bf 35} (2012) 29--98, [\href{http://xxx.lanl.gov/abs/1109.5515}{{\tt
  arXiv:1109.5515}}].

\bibitem{GomezCadenas:2010gs}
J.~J. Gomez-Cadenas, J.~Martin-Albo, M.~Sorel, P.~Ferrario, F.~Monrabal,
  et~al., {\it {Sense and sensitivity of double beta decay experiments}},  {\em
  JCAP} {\bf 1106} (2011) 007, [\href{http://xxx.lanl.gov/abs/1010.5112}{{\tt
  arXiv:1010.5112}}].

\bibitem{Auger:2012ar}
{\bf EXO} Collaboration, M.~Auger et~al., {\it {Search for Neutrinoless
  Double-Beta Decay in $^{136}$Xe with EXO-200}},  {\em Phys.Rev.Lett.} {\bf
  109} (2012) 032505, [\href{http://xxx.lanl.gov/abs/1205.5608}{{\tt
  arXiv:1205.5608}}].

\bibitem{Gando:2012zm}
{\bf KamLAND-Zen} Collaboration, A.~Gando et~al., {\it {Limit on Neutrinoless
  $\beta\beta$ Decay of Xe-136 from the First Phase of KamLAND-Zen and
  Comparison with the Positive Claim in Ge-76}},
  \href{http://xxx.lanl.gov/abs/1211.3863}{{\tt arXiv:1211.3863}}.

\bibitem{KlapdorKleingrothaus:2001ke}
H.~Klapdor-Kleingrothaus, A.~Dietz, H.~Harney, and I.~Krivosheina, {\it
  {Evidence for neutrinoless double beta decay}},  {\em Mod. Phys. Lett.} {\bf
  A16} (2001) 2409--2420, [\href{http://xxx.lanl.gov/abs/hep-ph/0201231}{{\tt
  hep-ph/0201231}}].

\bibitem{Bergstrom:2012nx}
J.~Bergstrom, {\it {Combining and comparing neutrinoless double beta decay
  experiments using different nuclei}},
  \href{http://xxx.lanl.gov/abs/1212.4484}{{\tt arXiv:1212.4484}}.

\bibitem{Alvarez:2012haa}
{\bf NEXT} Collaboration, V.~Alvarez et~al., {\it {NEXT-100 Technical Design
  Report (TDR): Executive Summary}},  {\em JINST} {\bf 7} (2012) T06001,
  [\href{http://xxx.lanl.gov/abs/1202.0721}{{\tt arXiv:1202.0721}}].

\bibitem{Alvarez:2012as}
V.~Alvarez et~al., {\it {Radiopurity control in the NEXT-100 double beta decay
  experiment: procedures and initial measurements}},
  \href{http://xxx.lanl.gov/abs/1211.3961}{{\tt arXiv:1211.3961}}.

\bibitem{Alvarez:2012hh}
{\bf NEXT} Collaboration, V.~Alvarez et~al., {\it {Near-Intrinsic Energy
  Resolution for 30 to 662 keV Gamma Rays in a High Pressure Xenon
  Electroluminescent TPC}},  \href{http://xxx.lanl.gov/abs/1211.4474}{{\tt
  arXiv:1211.4474}}.

\bibitem{Alvarez:2012nd}
{\bf NEXT} Collaboration, V.~\'Alvarez et~al., {\it {Initial results of
  NEXT-DEMO, a large-scale prototype of the NEXT-100 experiment}},
  \href{http://xxx.lanl.gov/abs/1211.4838}{{\tt arXiv:1211.4838}}.

\bibitem{Nygren:2009zz}
D.~Nygren, {\it {High-pressure xenon gas electroluminescent TPC for 0nu beta
  beta-decay search}},  {\em Nucl.Instrum.Meth.} {\bf A603} (2009) 337--348.

\bibitem{GomezCadenas:2012jv}
J.~Gomez-Cadenas, J.~Martin-Albo, and F.~Monrabal, {\it {NEXT, high-pressure
  xenon gas experiments for ultimate sensitivity to Majorana neutrinos}},  {\em
  JINST} {\bf 7} (2012) C11007, [\href{http://xxx.lanl.gov/abs/1210.0341}{{\tt
  arXiv:1210.0341}}].

\bibitem{WMAP}
C.~Bennett, D.~Larson, J.~Weiland, N.~Jarosik, G.~Hinshaw, et~al., {\it
  {Nine-Year Wilkinson Microwave Anisotropy Probe (WMAP) Observations: Final
  Maps and Results}},  \href{http://xxx.lanl.gov/abs/1212.5225}{{\tt
  arXiv:1212.5225}}.

\bibitem{BAO1}
L.~Anderson, E.~Aubourg, S.~Bailey, D.~Bizyaev, M.~Blanton, et~al., {\it {The
  clustering of galaxies in the SDSS-III Baryon Oscillation Spectroscopic
  Survey: Baryon Acoustic Oscillations in the Data Release 9 Spectroscopic
  Galaxy Sample}},  \href{http://xxx.lanl.gov/abs/1203.6594}{{\tt
  arXiv:1203.6594}}.

\bibitem{BAO2}
C.~Blake, E.~Kazin, F.~Beutler, T.~Davis, D.~Parkinson, et~al., {\it {The
  WiggleZ Dark Energy Survey: mapping the distance-redshift relation with
  baryon acoustic oscillations}},  {\em Mon.Not.Roy.Astron.Soc.} {\bf 418}
  (2011) 1707--1724, [\href{http://xxx.lanl.gov/abs/1108.2635}{{\tt
  arXiv:1108.2635}}].

\bibitem{BAO3}
N.~Padmanabhan, X.~Xu, D.~J. Eisenstein, R.~Scalzo, A.~J. Cuesta, et~al., {\it
  {A 2
  Methods and Application to the Sloan Digital Sky Survey}},
  \href{http://xxx.lanl.gov/abs/1202.0090}{{\tt arXiv:1202.0090}}.

\bibitem{H01}
A.~G. Riess, L.~Macri, S.~Casertano, H.~Lampeitl, H.~C. Ferguson, et~al., {\it
  {A 3
  Telescope and Wide Field Camera 3}},  {\em Astrophys.J.} {\bf 730} (2011)
  119, [\href{http://xxx.lanl.gov/abs/1103.2976}{{\tt arXiv:1103.2976}}].

\bibitem{H02}
W.~L. Freedman, B.~F. Madore, V.~Scowcroft, C.~Burns, A.~Monson, et~al., {\it
  {Carnegie Hubble Program: A Mid-Infrared Calibration of the Hubble
  Constant}},  {\em Astrophys.J.} {\bf 758} (2012) 24,
  [\href{http://xxx.lanl.gov/abs/1208.3281}{{\tt arXiv:1208.3281}}].

\bibitem{clusters}
B.~Benson, T.~de~Haan, J.~Dudley, C.~Reichardt, K.~Aird, et~al., {\it
  {Cosmological Constraints from Sunyaev-Zel'dovich-Selected Clusters with
  X-ray Observations in the First 178 Square Degrees of the South Pole
  Telescope Survey}},  \href{http://xxx.lanl.gov/abs/1112.5435}{{\tt
  arXiv:1112.5435}}.

\bibitem{Giusarma1}
E.~Giusarma, R.~de~Putter, and O.~Mena, {\it {Testing standard and non-standard
  neutrino physics with cosmological data}},
  \href{http://xxx.lanl.gov/abs/1211.2154}{{\tt arXiv:1211.2154}}.

\bibitem{Giusarma2}
G.-B. Zhao, S.~Saito, W.~J. Percival, A.~J. Ross, F.~Montesano, et~al., {\it
  {The clustering of galaxies in the SDSS-III Baryon Oscillation Spectroscopic
  Survey: weighing the neutrino mass using the galaxy power spectrum of the
  CMASS sample}},  \href{http://xxx.lanl.gov/abs/1211.3741}{{\tt
  arXiv:1211.3741}}.

\bibitem{Story}
K.~Story, C.~Reichardt, Z.~Hou, R.~Keisler, K.~Aird, et~al., {\it {A
  Measurement of the Cosmic Microwave Background Damping Tail from the
  2500-square-degree SPT-SZ survey}},
  \href{http://xxx.lanl.gov/abs/1210.7231}{{\tt arXiv:1210.7231}}.

\bibitem{Bond:1997wr}
J.~Bond, G.~Efstathiou, and M.~Tegmark, {\it {Forecasting cosmic parameter
  errors from microwave background anisotropy experiments}},  {\em
  Mon.Not.Roy.Astron.Soc.} {\bf 291} (1997) L33--L41,
  [\href{http://xxx.lanl.gov/abs/astro-ph/9702100}{{\tt astro-ph/9702100}}].

\bibitem{Zaldarriaga:1997ch}
M.~Zaldarriaga, D.~N. Spergel, and U.~Seljak, {\it {Microwave background
  constraints on cosmological parameters}},  {\em Astrophys.J.} {\bf 488}
  (1997) 1--13, [\href{http://xxx.lanl.gov/abs/astro-ph/9702157}{{\tt
  astro-ph/9702157}}].

\bibitem{GonzalezGarcia}
M.~Gonzalez-Garcia, M.~Maltoni, J.~Salvado, and T.~Schwetz, {\it {Global fit to
  three neutrino mixing: critical look at present precision}},  {\em JHEP} {\bf
  1212} (2012) 123, [\href{http://xxx.lanl.gov/abs/1209.3023}{{\tt
  arXiv:1209.3023}}].

\bibitem{GonzalezGarcia:2010un}
M.~Gonzalez-Garcia, M.~Maltoni, and J.~Salvado, {\it {Robust Cosmological
  Bounds on Neutrinos and their Combination with Oscillation Results}},  {\em
  JHEP} {\bf 1008} (2010) 117, [\href{http://xxx.lanl.gov/abs/1006.3795}{{\tt
  arXiv:1006.3795}}].

\bibitem{Cremonesi:2012av}
O.~Cremonesi, {\it {Experimental searches of neutrinoless double beta decay}},
  \href{http://xxx.lanl.gov/abs/1212.4885}{{\tt arXiv:1212.4885}}.

\bibitem{Sarazin:2012ct}
X.~Sarazin, {\it {Review of double beta experiments}},
  \href{http://xxx.lanl.gov/abs/1210.7666}{{\tt arXiv:1210.7666}}.

\bibitem{Zuber:2006hv}
K.~Zuber, {\it {Neutrinoless double beta decay experiments}},  {\em Acta Phys.\
  Polon.} {\bf B37} (2006) 1905--1921,
  [\href{http://xxx.lanl.gov/abs/nucl-ex/0610007}{{\tt nucl-ex/0610007}}].

\bibitem{Cattadori:2012fy}
C.~M. Cattadori, {\it {GERDA status report: Results from commissioning}},  {\em
  J.Phys.Conf.Ser.} {\bf 375} (2012) 042008.

\bibitem{Wilkerson:2012ga}
J.~Wilkerson, E.~Aguayo, F.~Avignone, H.~Back, A.~Barabash, et~al., {\it {The
  MAJORANA demonstrator: A search for neutrinoless double-beta decay of
  germanium-76}},  {\em J.Phys.Conf.Ser.} {\bf 375} (2012) 042010.

\bibitem{Gorla:2012gd}
{\bf CUORE} Collaboration, P.~Gorla, {\it {The CUORE experiment: Status and
  prospects}},  {\em J.Phys.Conf.Ser.} {\bf 375} (2012) 042013.

\bibitem{Auger:2012gs}
M.~Auger, D.~Auty, P.~Barbeau, L.~Bartoszek, E.~Baussan, et~al., {\it {The
  EXO-200 detector, part I: Detector design and construction}},  {\em JINST}
  {\bf 7} (2012) P05010, [\href{http://xxx.lanl.gov/abs/1202.2192}{{\tt
  arXiv:1202.2192}}].

\bibitem{Gando:2012jr}
{\bf KamLAND-Zen} Collaboration, A.~Gando et~al., {\it {First result from
  KamLAND-Zen: Double beta decay with $^{136}$Xe}},
  \href{http://xxx.lanl.gov/abs/1205.6130}{{\tt arXiv:1205.6130}}.

\bibitem{KamLANDZen:2012aa}
{\bf KamLAND-Zen} Collaboration, A.~Gando et~al., {\it {Measurement of the
  double-$\beta$ decay half-life of $^{136}$Xe with the KamLAND-Zen
  experiment}},  {\em Phys.Rev.} {\bf C85} (2012) 045504,
  [\href{http://xxx.lanl.gov/abs/1201.4664}{{\tt arXiv:1201.4664}}].

\bibitem{Ackerman:2011gz}
{\bf EXO-200} Collaboration, N.~Ackerman et~al., {\it {Observation of
  Two-Neutrino Double-Beta Decay in $^{136}$Xe with EXO-200}},  {\em
  Phys.Rev.Lett.} {\bf 107} (2011) 212501,
  [\href{http://xxx.lanl.gov/abs/1108.4193}{{\tt arXiv:1108.4193}}].

\bibitem{Abe:2009aa}
{\bf KamLAND Collaboration} Collaboration, S.~Abe et~al., {\it {Production of
  Radioactive Isotopes through Cosmic Muon Spallation in KamLAND}},  {\em
  Phys.Rev.} {\bf C81} (2010) 025807,
  [\href{http://xxx.lanl.gov/abs/0907.0066}{{\tt arXiv:0907.0066}}].

\bibitem{Conti:2003av}
{\bf EXO Collaboration} Collaboration, E.~Conti et~al., {\it {Correlated
  fluctuations between luminescence and ionization in liquid xenon}},  {\em
  Phys.Rev.} {\bf B68} (2003) 054201,
  [\href{http://xxx.lanl.gov/abs/hep-ex/0303008}{{\tt hep-ex/0303008}}].

\bibitem{MartinAlbo:2013xx}
J.~Martin-Albo and J.~J. Gomez-Cadenas, {\it {Status and physics potential of
  NEXT-100}},  {\em J.\ Phys.\ Conf.\ Ser.} (2013)
  [\href{http://xxx.lanl.gov/abs/1301.xxxx}{{\tt arXiv:1301.xxxx}}].

\bibitem{Kozlov:2011}
{\bf KamLAND-Zen} Collaboration, A.~Kozlov, {\it Status of the kamland-zen
  experiment},  in {\em {12th International Conference on Topics in
  Astroparticle and Underground Physics}}, Munich, 2011.

\end{thebibliography}\endgroup


\end{document}